\documentclass[prl,showpacs,floatfix,twocolumn,superscriptaddress]{revtex4}
\usepackage{graphicx}
\usepackage{dcolumn}
\usepackage{float}

\begin{document}

\bibliographystyle{apsrev}

\preprint{Draft, not for distribution} 
%
%
\title{Charge transfer in the high dielectric constant materials 
\boldmath CaCu$_3$Ti$_4$O$_{12}$ and CdCu$_3$Ti$_4$O$_{12}$ \unboldmath}
\author{C.C.~Homes}
\email{homes@bnl.gov}%
\author{T.~Vogt}
\author{S.M.~Shapiro}
\author{W.~Si}
\affiliation{Department of Physics, Brookhaven National Laboratory, Upton, NY 
11973}%
%
%
\author{S.~Wakimoto}
\altaffiliation{Present address: Department of Physics, University of Toronto, 
60 St.~George St., Toronto, Ontario, Canada, M5S 1A7.}%
\affiliation{Department of Physics, Brookhaven National Laboratory, Upton, NY 
11973}%
\affiliation{Department of Physics, Massachusetts Institute of Technology, 
Cambridge, MA 02139-4307 USA.}%
%
%
\author{M.A.~Subramanian}
\affiliation{DuPont Central Research \& Development Experimental Station, 
Wilmington, DE 19880-0328}%
%
%
\date{August 16, 2002}

%
%
\begin{abstract}
The cubic perovskite-related ceramic CaCu$_3$Ti$_4$O$_{12}$ has a very high 
static dielectric constant $\varepsilon_0\gtrsim 10\,000$ at room temperature 
(RT), which drops to about 100 below $\simeq 100$~K. Substituting Cd for Ca 
reduces the RT value of $\varepsilon_0$ by over an order of magnitude.  The 
large $\varepsilon_0$ may be due to an internal barrier layer capacitance 
(IBLC) effect. Infrared optical properties show a low-frequency mode that 
increases dramatically in strength at low temperature, suggesting a change in 
the effective charges and a breakdown of the IBLC model due to a 
semiconductor-to-insulator transition. 
\end{abstract}
%
%
\maketitle 

\begin{figure}[b]
\vspace*{-0.8cm}%
\centerline{\includegraphics[width=3.2in]{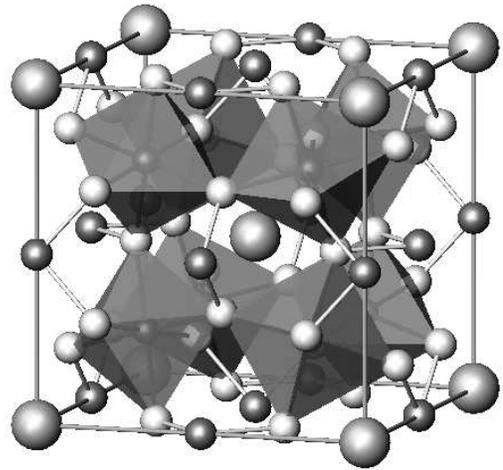}}%
\vspace*{-1.0cm}%
\caption{The unit cell of body-centered cubic CaCu$_3$Ti$_4$O$_{12}$ in the 
$Im\bar{3}$ space group.  The Ti atoms sit at the center of the TiO$_6$ 
octahedra, with bridging Cu atoms bonded to the oxygens, and large Ca atoms 
the corners and center of the unit cell.}%
\label{unitcell} 
\end{figure}

%
%
High dielectric constant materials find numerous technological applications.  
In the case of memory devices based on capacitive components, such as static 
and dynamic random access memories, the static dielectric constant 
$\varepsilon_0$ will ultimately decide the level of miniaturization.  The 
dielectric constant of a material is related to the polarizability $\alpha$, in 
particular the dipole polarizability (an atomic property), which arises from 
structures with a permanent electric dipole which can change orientation in an 
applied electric field.  These two quantities are linked through the 
Clausius-Mossotti relation.
%
%
In insulators $\varepsilon_0 > 0$; materials with a dielectric constant greater 
than that of silicon nitride ($\varepsilon_0 > 7$) are classified as ``high 
dielectric constant'' materials.  In general, a value of $\varepsilon_0$ above 
1000 is related to either a ferroelectric which exhibits a dipole moment in the 
absence of an external electric field, or a relaxor characterized by a 
ferroelectric response under high electric fields at low temperature, but no 
macroscopic spontaneous polarization. However, both classes of materials show a 
peak in $\varepsilon_0$ as a function of temperature, which is undesirable for 
many applications.  The body centered cubic (bcc) perovskite-related material 
CaCu$_3$Ti$_4$O$_{12}$ shown in Fig.~\ref{unitcell} has recently attracted a 
great deal of attention due to its extremely high value for the static 
dielectric constant $\varepsilon_0 \sim 10^4$ measured in ceramics in the radio 
frequency (kHZ) region \cite{subramanian00,ramirez00,homes01}, and was found to 
be practically constant in the $100-600$~K region.  Both properties are 
important for device implementation \cite{singh99,kim01}.  However, 
$\varepsilon_0$ displays a 100-fold reduction below $\sim 100$~K, without any 
detectable change in of the long-range crystallographic structure when probed 
by high-resolution x-ray \cite{ramirez00} and neutron powder diffraction 
\cite{subramanian00}.  This contrasts with known ferroelectrics, which 
structurally distort because of soft-mode condensation 
\cite{zeng99,transitions}. The substitution of Ca with Cd, results in a 
material with a similar temperature dependence, but a much lower dielectric 
constant, $\varepsilon_0\sim 500$ in a ceramic \cite{subramanian00}.  Concerns 
have recently been raised that the large values for $\varepsilon_0$ is purely 
an extrinsic effect due to Maxwell-Wagner-type depletion layers at sample 
contacts or at grain boundaries \cite{lunkenheimer02}.  However, recent 
measurements have been performed where the contacts were separated from the 
sample through the use of a thin aluminum oxide buffer layer; this showned that 
$\varepsilon_0 \sim 10^4$ for CaCu$_3$Ti$_4$O$_{12}$ \cite{artprivate}, ruling 
out contact contributions as the sole source of the large $\varepsilon_0$.  
Moreover, it is not clear why the substitution of Cd for Ca results in such a 
dramatically lower value for $\varepsilon_0$.  In this letter, we present 
optical results which offer insights as to the origin of the large 
$\varepsilon_0$ in CaCu$_3$Ti$_4$O$_{12}$ and its rapid decrease below $\simeq 
100$~K, as well as a possible explanation as to why $\varepsilon_0$ is much 
smaller in CdCu$_3$Ti$_4$O$_{12}$. 

%
%

%
%
Single crystals of CaCu$_3$Ti$_4$O$_{12}$ were grown by the traveling-solvent 
floating-zone method with an image furnace using a technique that has been 
described in detail elsewhere \cite{homes01}.
%
%
The ceramic compounds were prepared by conventional solid state reaction using 
starting oxides (CaO or CdO, TiO$_2$ and CuO) with a purity of 99.9\% or 
higher.  The mixed powder taken in stoichiometric ratio was calcined at 
850$^\circ$C for 8~hours.  The calcined powder was reground and pressed into 
disks and sintered in sealed gold tubes at 1000$^\circ$C for 20~hours.  X-ray 
diffraction data showed the materials are a single-phase.  The {\it 
A}Cu$_3$Ti$_4$O$_{12}$ family of compounds has been know for some time 
\cite{bochu79}, and their structures have been determined (Fig.~1). 

%
%
\begin{figure}[t]
\vspace*{-0.4cm}%
\centerline{\includegraphics[width=3.9in]{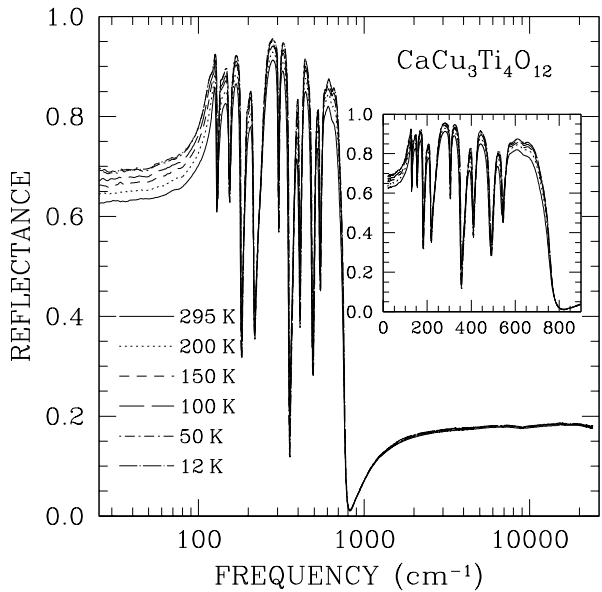}}%
\vspace*{-1.4cm}%
\caption{The temperature dependence of the reflectance of 
CaCu$_3$Ti$_4$O$_{12}$ from $\approx 20$ to $23\,000$~cm$^{-1}$.  The 
reflectance is typical for that of an insulator.  Above highest longitudinal  
optic phonon frequency ($\approx 700$~cm$^{-1}$), the reflectance becomes flat 
and featureless to the highest measured frequency, indicating that the gap edge 
has not yet been encountered (i.e., $2\Delta\gtrsim 3$~eV).  Inset: The 
low-frequency reflectance.}%
\label{reflec} 
\end{figure}

%
%
The temperature dependent reflectance of polished CaCu$_3$Ti$_4$O$_{12}$ 
(crystal) and  CdCu$_3$Ti$_4$O$_{12}$ (ceramic) have been measured over a wide 
range ($\approx 20$ to $23\,000$~cm$^{-1}$) using an overcoating technique 
\cite{homes93}.  In practice, the reflectance of ceramics and crystals is 
nearly identical.  The reflectance of the Ca material shown in 
Fig.~\ref{reflec} is typical of a non-metallic system.  The sharp features in 
the reflectance are due to the unscreened infrared active lattice vibrations; 
above the highest observed phonon frequency the reflectance is flat and 
featureless up to the highest measured frequency, suggesting that the optical 
value for the optical gap $2\Delta \gtrsim 3$~eV. 
%
%
The optical properties have been determined from a Kramers-Kronig analysis of 
the reflectance, which requires extrapolations at high and low frequencies.  At 
low frequency, the reflectance was assumed to be constant below the lowest 
measured frequency for $\omega\rightarrow 0$, while at high frequency the 
reflectance was assumed to be constant above the highest measured point to 
$2\times 10^5$~cm$^{-1}$, above which a free-electron approximation 
($R\propto\omega^{-4}$) was assumed. 
 
%
%
\begin{figure}
\vspace*{-0.4cm}%
\centerline{\includegraphics[width=3.9in]{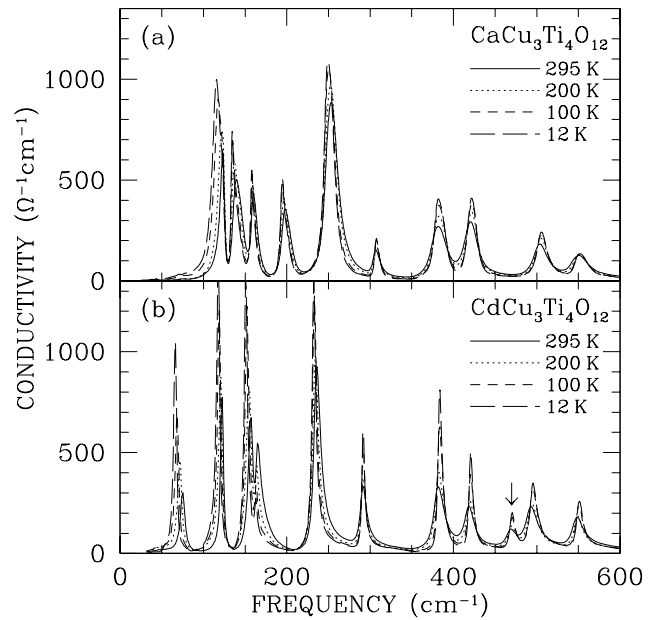}}%
\vspace*{-1.0cm}%
\caption{(a) The temperature dependent optical conductivity of 
CaCu$_3$Ti$_4$O$_{12}$. The low-frequency infrared-active lattice modes soften 
and show an anomalous increase in oscillator strength with decreasing 
temperature. 
(b) The temperature dependent optical conductivity of CdCu$_3$Ti$_4$O$_{12}$. 
The low-frequency modes show the same anomalous increase in strength with 
decreasing frequency, and in addition, many of the vibrations in this material 
are narrower at low temperature.  Note the unusual appearance of a new mode at 
$\sim 470$~cm$^{-1}$ (arrow) in the Cd material. (The vibrational parameters 
are tabulated in Table~I.)} %
\label{conduc} 
\end{figure}
 
%
%
The real part of the optical conductivity, derived from the imaginary part of 
the dielectric function $\sigma_1=\omega\varepsilon_2/4\pi$, is shown for 
CaCu$_3$Ti$_4$O$_{12}$ in Fig.~\ref{conduc}(a) in the low frequency region. As 
with the reflectance, the conductivity is characteristic of a semiconductor or 
insulator [$\sigma_{dc}\equiv \sigma_1(\omega\rightarrow 0)\approx 0$], and is 
dominated by the lattice modes.   The low-frequency modes display an anomalous 
increase in oscillator strength at low temperature. The  optical conductivity 
of CdCu$_3$Ti$_4$O$_{12}$ is shown in Fig.~\ref{conduc}(b).  The low frequency 
vibrations in the Cd material have the same anomalous increase in strength at 
low temperature.   In addition, a new mode is clearly observed at $\approx 
470$~cm$^{-1}$; this mode is curiously absent in the Ca material 
\cite{homes01,he01}, and will be discussed in more detail in a future work.  
Below $\approx 300$~cm$^{-1}$, there are some significant downward frequency 
shifts with doping with respect to the Ca material, indicating that Cd (Ca) 
plays a significant role in these vibrations.  
The infrared active modes have been fitted using the complex dielectric 
function $\tilde\varepsilon(\omega) = \varepsilon_1(\omega) + 
i\varepsilon_2(\omega)$ for Lorentz oscillators 
\begin{equation}
  \tilde\varepsilon(\omega) = \varepsilon_\infty+\sum_j {{\omega_{p,j}^2} \over 
  {\omega_j^2-\omega^2-i\omega\gamma_j}},
\end{equation}
where $\omega_j$, $\gamma_j$ and $\omega_{p,j}$ are the frequency, width and 
effective plasma frequency of the $j$th vibration; $\varepsilon_\infty$ is the 
core contribution to the dielectric function at high frequencies.  The results 
of the fits to the conductivity are shown in Table~I at 10 and 295~K for both 
materials.

%
%
%
\begin{table}[t]
\caption{The phonon parameters for Lorentzian fits to the conductivity of 
CaCu$_3$Ti$_4$O$_{12}$ and CdCu$_3$Ti$_4$O$_{12}$ at 10 and 295~K, where 
$\omega_j$, $\gamma_j$ and $\omega_{p,j}$ are the frequency, width and 
effective plasma frequency of the $j$th vibration. All units are in cm$^{-1}$.} 
\begin{ruledtabular}
\begin{tabular}{ccccccccccccc}
 \multicolumn{6}{c}{CaCu$_3$Ti$_4$O$_{12}^a$} & &
 \multicolumn{6}{c}{CdCu$_3$Ti$_4$O$_{12}^b$} \\
 \multicolumn{3}{c}{295~K} & \multicolumn{3}{c}{10~K} & & 
 \multicolumn{3}{c}{295~K} & \multicolumn{3}{c}{10~K} \\
  $\omega_j$ & $\gamma_j$ & $\omega_{p,j}$ & 
  $\omega_j$ & $\gamma_j$ & $\omega_{p,j}$ & & 
  $\omega_j$ & $\gamma_j$ & $\omega_{p,j}$ & 
  $\omega_j$ & $\gamma_j$ & $\omega_{p,j}$ \\  
  \cline{1-3} \cline{4-6} \cline {8-10} \cline{11-13}
  552 & 27  & 435 & 554 & 18  & 376 & & 550 & 26  & 484 & 552 & 13  & 421 \\
  504 & 19  & 445 & 506 & 14  & 454 & & 494 & 20  & 500 & 497 & 13  & 502 \\
  --- & --- & --- & --- & --- & --- & & 468 & 13  & 250 & 471 & 7.1 & 277 \\
  421 & 19  & 553 & 422 & 11  & 529 & & 419 & 17  & 450 & 421 & 6.1 & 421 \\
  382 & 18  & 535 & 383 & 13  & 560 & & 383 & 15  & 533 & 384 & 6.0 & 534 \\
  308 & 8.7 & 255 & 308 & 5.2 & 246 & & 292 & 13  & 471 & 292 & 4.8 & 410 \\
  254 & 16  & 913 & 251 & 12  & 916 & & 237 & 10  & 745 & 233 & 5.8 & 681 \\
  199 & 9.9 & 456 & 195 & 7.1 & 448 & & 166 & 13  & 575 & 163 & 8.0 & 291 \\
  161 & 7.8 & 423 & 159 & 5.2 & 388 & & 156 & 5.6 & 433 & 151 & 4.7 & 619 \\
  141 & 11  & 562 & 136 & 6.1 & 453 & & 122 & 4.0 & 434 & 118 & 4.6 & 636 \\
  122 & 6.0 & 464 & 116 & 12  & 851 & &  75 & 6.9 & 348 &  66 & 4.3 & 523 \\
\end{tabular}
\end{ruledtabular}
\footnotetext[1] {Single crystal and ceramic samples yield similar results.}%
\footnotetext[2] {Ceramic results only.} 
\end{table}

\begin{figure}[t]
\vspace*{-0.4cm}%
\centerline{\includegraphics[width=3.8in]{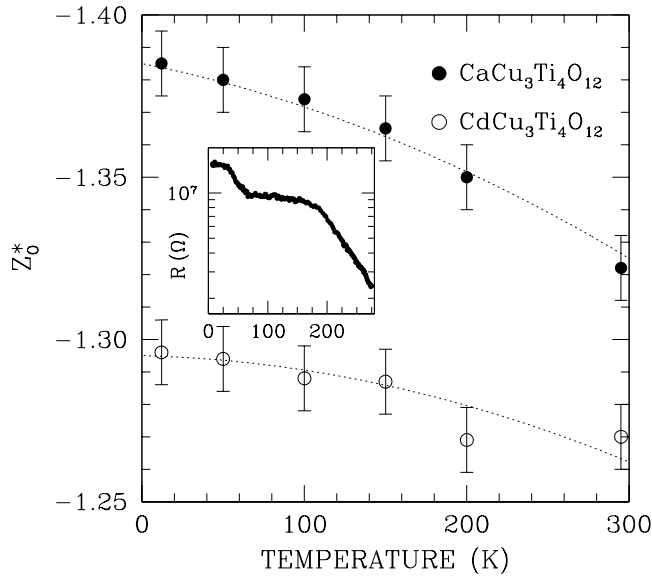}}%
\vspace*{-1.0cm}%
\caption{The temperature dependence of the deduced values Born effective charge 
per oxygen atom ($Z_{\rm O}^\ast$) in CaCu$_3$Ti$_4$O$_{12}$ and 
CdCu$_3$Ti$_4$O$_{12}$.  The dotted lines are drawn as a guide to the eye.  
Note that the value for $Z_{\rm O}^\ast$ in the Cd material is lower for the Ca 
material, and that the $Z_{\rm O}^\ast$ increases by more than 5\% in 
CaCu$_3$Ti$_4$O$_{12}$, compared to an increase of less 
than 2\% in CdCu$_3$Ti$_4$O$_{12}$.  
Inset: The temperature dependence of the resistance of a thin film of 
CaCu$_3$Ti$_4$O$_{12}$.  The resistance increases with decreasing temperature, 
but there is a discontinuity close to 100~K.}%
\label{charges} 
\end{figure}

%
%
The anomalous increase in oscillator strength of the low frequency modes is 
unusual, and has important consequences.  Optical sum rules provide a powerful 
tool with which to analyze the behavior of free carriers and bound excitations 
\cite{smith85}.  The partial conductivity sum rule for oscillators states that 
\cite{units} 
\begin{equation}
  {120\over \pi} \int_{\omega_a}^{\omega_b} \sigma_1(\omega)\,d\omega = 
  \omega_{p,j}^2 ,
\end{equation}
where the interval $\omega_a \rightarrow \omega_b$ is chosen so that the full 
spectral weight of the $j$th oscillator is captured.  In the absence of changes 
in the bonding or coordination, the mode may narrow with decreasing 
temperature, but the spectral weight (proportional to area under the peak, or 
$\omega_{p,j}^2$) should not change.  The dramatic increase in the oscillator 
strength of the low-frequency mode (Fig.~\ref{conduc} and Table I) is a clear 
violation of this sum rule, which in turn has implications for the distribution 
of charge within the unit cell.  Light couples to the induced dipole moments 
created by the atomic displacements associated with a normal mode --- if the 
Born effective charge per atom $Z^\ast$ is increasing, then the size of the 
induced dipole moment and the optical absorption will also increase.  For a 
material with $k$ atoms in the unit cell, the effective charge per atom can be 
defined as \cite{scott71} 
\begin{equation}
  {1\over\epsilon_\infty} \sum_j \omega_{p,j}^2 =
  {{4\pi} \over V_c} \sum_k {{(Z_k^\ast e)^2} \over {M_k}}
  \label{charge}
\end{equation}  
where $\sum_k Z_k^\ast=0$, $V_c$ is the unit cell volume, and $j$ and $k$ index 
the lattice modes and the atoms with mass $M_k$, respectively.  The fitted 
values of $\omega_{p,j}$ in Table~I indicate that there is an increase between 
room temperature and 10~K in the left side of Eq.~\ref{charge} of $\simeq 11$\% 
in CaCu$_3$Ti$_4$O$_{12}$, and $\simeq 4$\% in CdCu$_3$Ti$_4$O$_{12}$.  The 
increase in the left side of Eq.~\ref{charge} implies that the $Z_k^\ast$'s are 
increasing with decreasing temperature.  
%
%
In oxide materials, oxygen is often the lightest element, so that the summation 
is dropped and the change in the effective charge is associated purely with the 
oxygen (i.e., $Z_k^\ast \equiv Z_{\rm O}^\ast$).  While the presence of other 
light elements in CaCu$_3$Ti$_4$O$_{12}$ may complicate this approach, it is 
less of a problem in the Cd material.  The deduced values for $Z_{\rm O}^\ast$ 
are shown in Fig.~\ref{charges}, and illustrate a noticeably different behavior 
for the Ca and Cd materials.  While Cd is somewhat smaller ($\sim 16$\%) than 
the Ca cation, the main difference between these two materials is their 
electronegativity, or the ability of an atom to attract electrons to itself. 
The Pauling electronegativity for Ca and Cd are 1.0 and 1.7, respectively.  It 
is expected that an atom with a higher electron affinity will result in less 
charge transfer to the oxygen atoms, and the reduction in $Z_{\rm O}^\ast$ is 
precisely what is in fact observed in the Cd material. 
  
%
%

%
%
The large dielectric constant observed in these materials must ultimately be 
due to either intrinsic effects that arise from the properties of the material, 
or extrinsic effects, such as contact problems \cite{lunkenheimer02}.  However, 
the persistence of the large value of $\varepsilon_0$ with the addition of a 
buffer layer between the sample and the contact indicates that contact problems 
alone are not the source of the large dielectric constant.  On the other hand, 
the absence of a structural transition tends to rule out the possibility that 
the large $\varepsilon_0$ is the result of intrinsic effects, such as the 
displacements of Ca atoms or some distortion that involves the TiO$_6$ 
octahedra. 
%
%
A more compelling explanation comes from the observation that 
CaCu$_3$Ti$_4$O$_{12}$ is heavily twinned \cite{subramanian00}.  Recent results 
based on impedance spectroscopy on CaCu$_3$Ti$_4$O$_{12}$ ceramics indicate 
that these materials may be understood as being semiconducting regions 
separated by insulating barriers, and that the giant dielectric phenomena is 
attributed to an internal barrier layer capacitance (IBLC) effect 
\cite{sinclair02}.  The IBLC results in a large $\varepsilon_0$ that has the 
same Debye-like frequency dependence that has been observed in 
CaCu$_3$Ti$_4$O$_{12}$ \cite{ramirez00,homes01}.  However, one of the more 
puzzling aspects of CaCu$_3$Ti$_4$O$_{12}$, and to a lesser extent 
CdCu$_3$Ti$_4$O$_{12}$, is the rapid suppression of $\varepsilon_0$ at low 
temperature.  The large values for $\varepsilon_0$ have been shown to persist 
in thin films \cite{lin02}, and transport measurements of 
CaCu$_3$Ti$_4$O$_{12}$ thin films shown in the inset of Fig.~4 indicate that 
there is an anomaly in the resistance at $\sim 100$~K that is suggestive of a 
semiconductor-to-insulator (SI) transition; such a transition would result in 
the rapid expansion of the insulating domains in the IBLC picture and a 
commensurate reduction of $\varepsilon_0$.  Furthermore, the SI transition is 
consistent with the observation of the increasing degree of ionicity within the 
unit cell of CaCu$_3$Ti$_4$O$_{12}$.  The lower value of $\varepsilon_0$ in 
CdCu$_3$Ti$_4$O$_{12}$ suggests that the material is not as heavily twinned, 
which may in turn be related to the slightly lower values of $Z_{\rm O}^\ast$ 
in this material. 

%
%
In summary, the optical properties of CaCu$_3$Ti$_4$O$_{12}$ and 
CdCu$_3$Ti$_4$O$_{12}$ have been measured at a variety of temperatures.  A low 
frequency mode is observed to strengthen dramatically at low temperature, 
indicating that the Born effective charges are increasing in the unit cell. We 
propose that the large $\varepsilon_0$ may be due to extrinsic mechanisms such 
as the formation of boundary-layer capacitors, and that the rapid reduction of 
$\varepsilon_0$ at low temperature is due to an SI transition and the removal 
of an IBLC mechanism.  The CdCu$_3$Ti$_4$O$_{12}$ system may have a lower value 
of $\varepsilon_0$ due to a reduced degree of twinning. 

%
%
\begin{acknowledgments}
We are grateful to M.H.~Cohen, L.~He, J.B.~Neaton, A.W.~Sleight, M.~Strongin, 
J.J.~Tu and D.~Vanderbilt for many useful discussions. 
This work was supported by the Department of Energy under contract number 
DE-AC02-98CH10886. 
\end{acknowledgments}

%
%
%
\bibliography{cctioref}

\begin{thebibliography}{17}
\bibitem[{uni()}]{units}
\bibinfo{note}{The prefactor in the sum rule assumes that the units of the
  conductivity are in $\Omega^{-1}$cm$^{-1}$, and that the units of the
  effective plasma frequency for the oscillator are in cm$^{-1}$.}
\end{thebibliography}

%

\end{document}